\documentstyle[10pt,epsf]{article}

\advance \topmargin by -\headheight
\advance \topmargin by -\headsep

\evensidemargin \oddsidemargin
 \advance\hoffset by -3mm  
 \advance\voffset by  8mm  

\def\half{{{1\over 2}}}

\def\pr{Phys. Rev.}
\def\ap{Ann. Phys.}
\def\cmp{Comm. Math. Phys.}

\def\jp{J. Phys.}

\newcommand{\beq}{\begin{equation}}
\newcommand{\eeq}{\end{equation}}
\newcommand{\bear}{\begin{eqnarray}}
\newcommand{\eear}{\end{eqnarray}}
\newcommand{\sgn}{{\mathrm{sgn}}}

\newcommand{\A}{{\bf A}}
\newcommand{\rr}{{\bf r}}
\newcommand{\p}{{\bf p}}
\newcommand{\frnu}{{{\{}\nu{\}}}}
\newcommand{\e}{\mathrm{e}}
\newcommand{\im}{\mathrm{i}}
\newcommand{\ack}{{\bf Acknowledgements}}

\newfont{\addfont}{cmti7 scaled 1440}
\newfont{\headfontb}{cmbx10 scaled 1728}
\begin{document}
\begin{titlepage}
\begin{center} {\headfontb Aharonov-Bohm scattering on a cone}
\end{center}
\vskip 0.3truein
\begin{center}
{\Large{M. Alvarez}}
\end{center}
\vskip 0.2truein
\begin{center} {\addfont{Department of Physics,}}\\
{\addfont{University of Wales Swansea}}\\ {\addfont{Singleton Park, Swansea 
SA2 8PP, U.K.}} \\ {\tt{e-mail: pyma@swansea.ac.uk}}
\end{center}
\vskip 1truein
\begin{center}
\bf ABSTRACT
\end{center} 
The Aharonov-Bohm scattering amplitude is calculated in the context of planar gravity with localized sources which also 
carry a magnetic flux. These sources cause space-time to develop conical singularities at their location, thus introducing 
novel effects in the scattering of electrically charged particles. The behaviour of the wave function in the proximity of the 
classical scattering directions is analyzed by means of an asymptotic expansion previously introduced by the author. It is 
found that, in contrast with the Aharonov-Bohm effect in flat space, integer values of the numerical flux can produce 
observable effects.
\vskip2.5truecm
\leftline{SWAT-98-189  \hfill April 1998}
\leftline{physics/9804032}

\end{titlepage}
\setcounter{footnote}{0}

\section{Introduction}
The Aharonov-Bohm (-Ehrenberg-Siday) effect \cite{ehre, aha} is a well-known example of quantum scattering process 
without classical analog, and its importance is reflected in the large number of analyses that have appeared in the literature 
(see the monograph \cite{pes} for a thorough exposition and a long list of references). The idealized experiment that 
produces the Aharonov-Bohm (AB) effect in its simplest version requires little description: charged particles of non-zero 
mass propagate at right angles to an infinitely long straight solenoid that encloses a magnetic flux; the region
containing the magnetic field is inaccessible to the particles, in spite of which the magnetic flux inside the solenoid affects 
their propagation and an interference pattern appears that cannot be explained in classical physics. Among many references, 
\cite{berry,berry2} are particularly relevant and readable. Detailed descriptions of real experimental set-ups can be found in 
\cite{pes} but we shall not be concerned with the practical aspects of the problem. 

In this work we shall introduce a further complication in the problem: our flux tubes will have a non-vanishing mass density 
and will introduce gravitational effects in the motion of the scattered particles. In order to make the problem amenable we 
shall restrict ourselves to the simplest situation of an infinitely long and thin straight flux tube with constant magnetic flux 
and mass density. We shall sometimes call the idealized flux tube ``string". The advantage of the restriction just 
mentioned is that the gravitational field of the string can be described by general relativity in $2+1$ dimensions \cite{djt,dj}, 
as the third spatial dimension can be taken to be parallel to the flux and will decouple from the problem. In the absence of 
other sources of gravity, the space created by the string will be locally flat everywhere except at the location of the string, 
and globally will be a $2$-cone. The time dimension remains unaffected as the string is taken to be static and at rest. The 
charged particle will be assumed to be massive but light enough to have negligible effect on the gravitational field of the 
string.

Given the previous simplifications, the object of interest is the wave function and scattering amplitude of a charged test 
particle moving in the conical background created by the flux line, hence the title of this paper. Naturally the solution will be 
a combination of planar gravitational scattering and AB scattering. The pure gravitational scattering amplitude was 
calculated in \cite{djt,dj,js,acg} (see also \cite{dow}), and the pure AB scattering amplitude in \cite{aha,berry3}-\cite{yo} 
following different procedures. The definition of the scattering amplitude requires knowledge of the long-distance behaviour 
of the scattered wave function, and usually the limit $r\to\infty$, with $r$ the distance from the scattering center, can be 
taken safely. One of the peculiarities of our problem is that, as in the purely gravitational case, there are two distinguished 
scattering directions independent of the energy of the incident particle that we call ``classical" scattering directions, and the 
limits ``approaching a classical scattering direction" and ``$r\to\infty$" do not commute. If we take the long-distance limit 
first, the wave function develops singularities at the  classical scattering directions, and if we approach those directions first 
then the long-distance limit does not exist. Physically the problem is that at the classical scattering directions the splitting 
of the wave function into ``scattered" and ``incident" (or transmitted) parts is no longer meaningful; only the complete wave 
function is free from discontinuities or singularities \cite{acg,yo}. The scattering amplitude, therefore, cannot be defined at 
those directions. This phenomenon occurs in the AB effect at the forward direction and in scattering in planar gravity at two 
directions that depend on the mass of the scattering center only \cite{djt,dj}. 

The analysis of the wave function near the classical scattering directions is best performed in the approach of \cite{acg} for 
planar gravitational scattering and \cite{yo} for AB scattering, and here the same approach will be used to solve the more 
general case of the AB effect on a cone. This article is organized as follows. In Section 2 we shall review the non-relativistic 
propagator of a charged particle in presence of a massive magnetic flux line. A piece of the propagator is given as an integral 
that is calculated in Section 3, where the result is used to obtain an asymptotic expansion of the propagator. We find that 
gravitational effects split the wave function into two halves that propagate along the classical scattering directions, and the 
two halves carry opposite AB-like phases. The last Section contains a discussion of the results; we conclude that, in 
contrast with the pure AB effect, the integer part of the numerical flux can be measured by means of an (imaginary) 
interference experiment. 

\section{The quantum-mechanical propagator}
We are interested in the problem of a charged particle moving on a conical background created by a massive flux tube 
assumed to coincide with the $z$-axis. We recall that in this situation (quantum scattering of a test-particle by a static 
mass) the time-component of the metric does not play any role. A convenient characterization of this conical space is based 
on embedded coordinates \cite{dj}
\beq
(dl)^{2}=\alpha^{-2}(dr)^{2}+r^2(d\theta)^2,
\qquad\qquad -\pi\leq\theta\leq\pi,
\label{22}
\eeq
where $0\leq(1-\alpha)=4MG<1$ and $G$ is ``Newton's constant''. If we consider that the incoming charged particles 
approach perpendicularly the flux line, the scattering process is essentially two-dimensional. In this situation a possible 
choice of vector potential is
\beq
\A(\rr)={\Phi\over 2\pi}\nabla\theta={c\hbar\over e}\nu\nabla\theta,
\label{vectorpot}
\eeq
where $\Phi$ is the flux carried by the string, $\nu$ is the ``numerical flux'' defined as $\nu=e\Phi/2\pi\hbar c$ and $\theta$ is 
the polar angle of cylindrical coordinates described above. The Hamiltonian that defines the dynamics of
the system is
\begin{eqnarray*}
H=-{\hbar^2\over 2m}\left(\p-{e\over c}\A(\rr)\right)^ig_{ij}\left(\p-{e\over c}\A(\rr)\right)^j 
\end{eqnarray*}
with $\A(\rr)$ given by (\ref{vectorpot}) and the metric $g_{ij}$ corresponding to the line element (\ref{22}). Based on 
previous results \cite{acg,yo} we expect the wave function to exhibit two privileged scattering directions that depend 
only on the mass density of the flux line, and superimposed on each of these directions AB-like phases induced by the 
magnetic flux . As the separate cases of planar gravitational scattering and AB scattering are known, all we have to do is 
combine the propagators analyzed in the above mentioned references. The result can be given as a Bessel series \cite{dow}
\begin{eqnarray*}
G(r,\theta;r',\theta';\alpha;\nu;t)={m\over 2\pi \im\hbar t\alpha^2}\exp\left\{\im{m\over 2\hbar
t\alpha^2}(r^2+r'^2)\right\}\sum_{n=-\infty}^{\infty}\e^{\im n(\theta-\theta')}\,\e^{-\im{\pi\over 2\alpha}|\nu-n|}
\,J_{{1\over\alpha}|\nu-n|}\left({mrr'\over \hbar t\alpha^2}\right),
\end{eqnarray*}
or, after inserting the Schl\"afli representation for the Bessel functions \cite{jack,dj,js}, as the sum of a ``transmited" plus a 
``scattered" part, the last one being in integral form:
\bear
G_{\rm{tr}}(r,\theta;r',\theta';\alpha;\nu;t)&=&{m\over 2\pi \im\hbar t\alpha}
\,\e^{\im\nu\phi} \mathop{{\sum}'}_{n=-\infty}^{\infty} \exp\Big{\{}\im
{m\over 2\hbar t\alpha^2}\big[ r^2+r'^2-2rr'\cos \alpha(\theta-\theta'+2\pi n)
\big]\Big{\}}\,\e^{\im 2\pi n\nu}, \nonumber\\
G_{\rm{sc}}(r,\theta;r',\theta';\alpha;\nu;t)&=&-{m\over 4\pi^2\hbar t\alpha^2}
\exp\left\{\im{m\over2\hbar t\alpha^2}(r^2+r'^2)\right\}\,\e^{\im[\nu]\phi} \int\limits_{-\infty}
^{\infty}dy\,\exp\left\{\im{mrr'\over\hbar t\alpha^2}\cosh y +{1\over\alpha}\frnu y\right\}\nonumber\\ & &\times \left[ 
{\e^{-\im\frnu{\pi\over\alpha}}\over 1-\e^{-\im(\theta-\theta')+{1\over\alpha}(y-\im\pi)}}-{\e^{\im\frnu{\pi\over\alpha}}\over
1-\e^{-\im(\theta-\theta')+{1\over\alpha}(y+\im\pi)}}\right],
\label{green}
\eear
where the primed sum includes only $n$ such that $\alpha(\theta-\theta'+2\pi n)\in(-\pi, \pi)$. The symbols $[\nu]$ and $\{ 
\nu \}$ are the integer and fractional parts of the numerical flux $\nu$. The suffixes of the two parts of the propagator in 
(\ref{green}) indicate that one part corresponds to the transmitted and the other to the scattered waves, although this 
splitting of the total propagator should not be taken literally because, as we shall see, both parts are required to determine 
the wave function of the ``scattered" particle at the classical scattering directions. The propagator (\ref{green}) reduces to 
the AB propagator if $\alpha=1$ and to the Deser-Jackiw propagator \cite{dj} of planar gravitational scattering if $\nu=0$. As 
in \cite{acg,yo} the propagator (\ref{green}) leads to an integral than can be calculated by a saddle-point approximation for 
all scattered directions except the classical ones. The integral in question is
\beq
I(\rho,\phi,\alpha,\nu)=\int\limits_{-\infty}^{\infty}dy\,\exp\left\{\im\rho\cosh y+{1\over\alpha}\frnu y\right\} \left[{\e^{-\im\frnu 
{\pi\over\alpha}}\over 1-\e^{-\im\phi+{1\over\alpha}(y-\im\pi)}}-{\e^{\im\frnu 
{\pi\over\alpha}}\over 1-\e^{-\im\phi+{1\over\alpha}(y+\im\pi)}}\right].
\label{int}
\eeq
where we have defined the parameters $\rho$ and $\phi$ as
\begin{eqnarray*}
\rho&=&{mrr'\over \hbar t\alpha^2},\nonumber\\
\phi&=&\theta-\theta'. 
\end{eqnarray*}
The saddle-point calculation of (\ref{int}) requires the limit $\rho\to\infty$ to be taken in the exponential part of the integrand 
and then the integral is concentrated about $y\approx 0$ and is approximately Gaussian. This calculation has been done 
before \cite{jack,dj,js} and does not need to be repeated here. We simply quote the resulting expression 
for the scattered propagator:
\beq
G_{\rm{sc}}(r,\theta;r',\theta';\alpha;\nu;t)\sim \e^{\im{3\over 4}\pi}\sqrt{ m\over 8\pi^3\hbar t\alpha^2rr'} 
\exp\left\{\im{m\over2\hbar t\alpha^2}(r+r')^2\right\}\e^{\im[\nu]\phi}{ 
\e^{\im\phi}\sin{\frnu\pi\over\alpha}+\sin({\pi\over\alpha}(1-\frnu)) \over 
\cos\phi-\cos{\pi\over\alpha}}.
\label{saddle}
\eeq
We have used the notation $\sim$ to indicate that the equation is an asymptotic expansion for large $\rho$ \cite{erdelyi}. 
From this result we can immediately write down the scattering amplitude of a well-localized wave packet approaching the 
string from $\theta'=\pi$ and incident momentum $k$ (with the same result for plane-wave scattering):
\beq
f(k, \theta)\sim {1\over \sqrt{2\pi k}} \e^{\im[\nu]\theta} 
\e^{-\im\frnu\pi}\,{\e^{\im\theta}\sin{\frnu\pi\over\alpha}-\sin({\pi\over\alpha}(1-\frnu))\over \cos\theta+\cos{\pi\over\alpha}}.
\label{scatam1}
\eeq
This scattering amplitude reduces to the AB or to the planar gravitational case in the appropriate limits. Naturally the global 
phases included in (\ref{scatam1}) are irrelevant if we are interested in the scattering cross section $\sigma=|f|^2$ only. Let 
us now remember that the classical scattering directions in our problem correspond, if the incident angle is $\theta'=\pi$, to 
the following two values for the scattering angle $\theta$ \cite{djt,dj,js}:
\begin{eqnarray*}
\theta_{\pm}=\pm \left({\pi\over\alpha}-\pi\right).
\end{eqnarray*}
The fact that the scattering amplitude (\ref{scatam1}) diverges at these two directions is of course not due to any 
pathologies of the scattering process but rather to the fact that, around $\theta=\theta_{\pm}$, the saddle point 
approximation is not warranted. In the next section we shall follow the approach of \cite{acg,yo} to determine the wave 
function at the two directions $\theta_{\pm}$.

\section{Asymptotic expansions}
We will now develop an asymptotic expansion of the integral $I(\rho,\phi,\alpha,\nu)$ defined in (\ref{int}) that, unlike a mere 
saddle-point approximation, be applicable when the scattering angle $\theta$ is close or equal to the classical scattering 
directions $\theta_{\pm}$. To that end we write the integral (\ref{int}) as a sum of two terms 
\begin{eqnarray*}
I(\rho,\phi,\alpha,\nu)&=&\e^{-\im\frnu{\pi\over\alpha}}\int\limits_{-\infty}^{\infty}dy\,\exp\left\{\im\rho\cosh y 
+{1\over\alpha}\frnu y\right\}  {1\over 1-\e^{-\im\phi+{1\over\alpha}(y-\im\pi)}} \\& &-
\e^{\im\frnu{\pi\over\alpha}}\int\limits_{-\infty}^{\infty}dy\,\exp\left\{\im\rho\cosh y +{1\over\alpha}\frnu y\right\} 
{1\over 1-\e^{-\im\phi+{1\over\alpha}(y+\im\pi)}}  \\&\equiv& 
\e^{-\im\frnu{\pi\over\alpha}}I_1(\rho,\phi,\alpha,\nu)-\e^{\im\frnu{\pi\over\alpha}}I_2(\rho,\phi,\alpha,\nu),
\end{eqnarray*}
and we can consider one of the two integrals, say $I_1$, and treat the other one by analogy. Following the lines proposed in 
\cite{acg,yo} the integral $I_1$ can be expanded as a hypergeometric series that, as we shall see, has a finite discontinuity at 
the classical scattering directions: 
\beq
I_1(\rho,\phi,\alpha,\nu)\sim \e^{-\im{\pi\over 4}}\sqrt{2}\, \e^{\im\rho}\sum_{m=0}^{\infty} \rho^{\half-m}\im^m 
A_{2m}(\phi+{\pi\over\alpha},\frnu,\alpha)\,\Gamma(m-\half)\,F_2(1,-m+{3\over2}, \im a(\alpha\phi)\rho),
\label{expan}
\eeq
where the coefficients $A_{2m}(\phi+{\pi\over\alpha},\frnu,\alpha)$ and $a(\alpha\phi)$ are defined by the following 
relations
\begin{eqnarray*}
{\cos\eta(s)-\cos(\alpha\theta)\over 
1-\e^{\im({\eta(s)\over\alpha}-\theta)}}{\e^{\im{\frnu\over\alpha}\eta(s)}\over\cos{\eta(s)\over2}}&=&\sum_{m=0}^{\infty}
\e^{\im m{\pi\over4}}A_m(\theta,\frnu,\alpha)s^m,  \\
\eta(s)&=&2\arcsin\left(\e^{-\im{\pi\over4}}{1\over\sqrt{2}}s\right),\\
a(\alpha\phi)&=&\e^{-\im\pi}(1+\cos(\alpha\phi)).  
\end{eqnarray*}
These calculations are rather lengthy, although straightforward, and can be reproduced after following the explanations 
given in \cite{acg,yo}. Using these results we can consider the limit $a(\alpha\phi)\rho\to\infty$, which corresponds to long 
distances from the scattering center away from the classical scattering directions $\theta_{\pm}$. Although the actual 
calculations are a good illustration of the use of our expansion (\ref{expan}), the final result coincides with (\ref{saddle}) and 
(\ref{scatam1}) and therefore is of little relevance. The true interest of (\ref{expan}) is its good behaviour about 
$\theta=\theta_{\pm}$, where the total integral develops a finite discontinuity. If for example $\phi=-\pi/\alpha+\epsilon$ with 
$\epsilon$ a very small angle, we obtain
\begin{eqnarray*}
I(\rho,-{\pi\over\alpha}+\epsilon, \alpha, 
\nu)\sim-\sgn(\epsilon)\pi\alpha\,\im\,\e^{-\im\frnu{\pi\over\alpha}}\e^{\im\rho}+\e^{\im{\pi\over4}}\e^{\im\rho}
\sqrt{{2\pi\over\rho}}\left[ \e^{-\im\frnu{\pi\over\alpha}}\left( \half-\frnu\right) + {\e^{\im\frnu{\pi\over\alpha}}\over 
\e^{\im{2\pi\over\alpha}}-1}\right] +\ldots
\end{eqnarray*}
where the dots indicate subdominant terms in the large $\rho$ limit. The scattered part of the propagator is therefore
\begin{eqnarray*}
G_{\rm{sc}}(r,-{\pi\over\alpha}+\pi+\epsilon;r',\pi;\alpha;\nu;t)={m\over 2\pi \im\hbar t\alpha^2}\exp\left\{\im{m\over 2\hbar
t\alpha^2}(r+r')^2\right\}\,\left(-\sgn(\epsilon)\right)\half\alpha \e^{-\im\nu{\pi\over\alpha}}+\ldots.
\end{eqnarray*}
The transmited propagator can be easily calculated by means of its explicit expresion given in (\ref{green}) with the following 
result:
\begin{eqnarray*}
G_{\rm{tr}}(r,-{\pi\over\alpha}+\pi+\epsilon;r',\pi;\alpha;\nu;t)={m\over 2\pi \im\hbar t\alpha^2}\exp\left\{\im{m\over 2\hbar
t\alpha^2}(r+r')^2\right\}\,\left(1+\sgn(\epsilon)\right)\half\alpha \e^{-\im\nu{\pi\over\alpha}}+\ldots.
\end{eqnarray*}
Both discontinuities cancel out in the complete propagator and thus the complete wave function is finite at 
$\phi=-\pi/\alpha$. A similar result obtains at the other classical scattering direction $\phi=\pi/\alpha$, with different sign in 
the phase $\nu\pi/\alpha$. Clearly the complete propagators so obtained represent linear propagation of wave packets along 
the classical scattering directions.

\begin{figure}
\centerline{\hskip.4in\epsffile{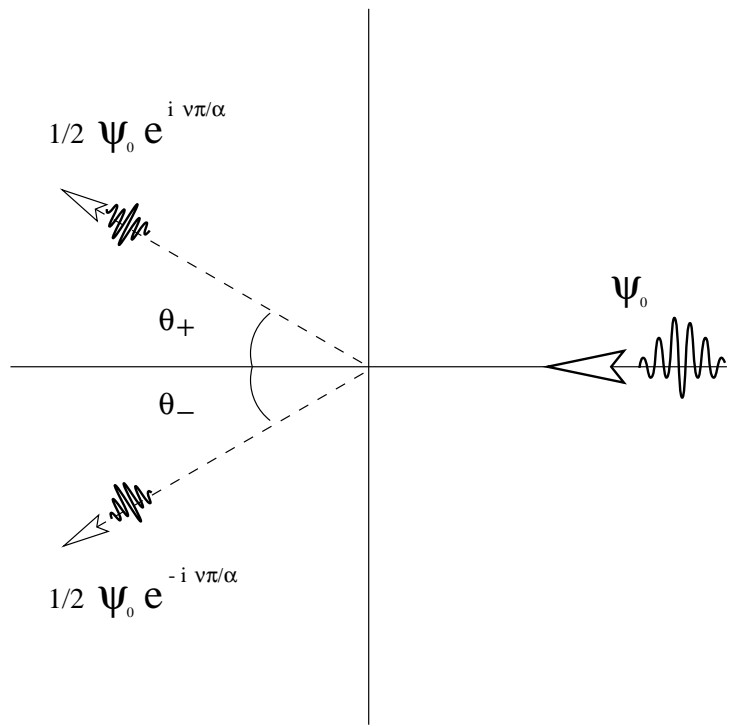}}
\caption{Aharonov-Bohm scattering of a wave packet on a cone}
\end{figure}

\section{Interpretation and conclusions}
The most interesting consequences of the propagators just obtained are that, at leading order in the long-distance limit, a 
wave packet approaching the magnetic tube will split into two halves that propagate along the classical scattering 
directions as in the case of pure gravitational scattering \cite{acg}, and that the two halves carry opposite phases that 
depend on the combined parameter $\nu/\alpha$ only. The situation is schematically represented in Fig.~1, and agrees with  
the purely gravitational or AB cases in the appropriate limits.

The fact that the opposite phases of the two emerging wave packets depend on the whole numerical flux $\nu$ opens the 
possibility of measuring $\nu$ by means of an interference experiment. That imaginary experiment would consist in reuniting 
both diverging wave packets and allowing them to interfere on a flat screen perpendicular to the incident beam; the resulting 
interference pattern will show bands whose shift from the centered position depends on $\nu/\alpha$. As the parameter 
$\alpha$ can be measured from the scattering angle, $\nu$ can in principle be determined. This contrasts with the purely AB 
case, where only the fractional part of the flux can be measured. 
\vskip .5in
\ack
\vskip .25in
This research has been supported by the Engineering and Physical Sciences Research Council (EPSRC) of the United 
Kingdom.


\end{document}